\documentclass[10pt,conference]{IEEEtran}

\usepackage{cite}
\usepackage{amsmath,amssymb,amsfonts}
\usepackage{algorithmic}
\usepackage{graphicx}
\usepackage{textcomp}
\usepackage[table,xcdraw]{xcolor}
\usepackage{orcidlink}
\usepackage{arydshln}
\usepackage{cleveref} 
\usepackage{enumitem}
\usepackage{xspace}
\usepackage{listings}
\usepackage{xargs}  
\usepackage[table]{xcolor}
\usepackage[colorinlistoftodos,prependcaption,textsize=small]{todonotes}
\usepackage[many]{tcolorbox}    	
\usepackage{booktabs}
\usepackage{multirow}
\usepackage{array}
\usepackage{caption}
\usepackage{ragged2e}
\usepackage{float}

\begin{document}

\newcommand{\eg}{\emph{e.g.},\xspace}
\newcommand{\ie}{\emph{i.e.},\xspace}
\newcommand{\wrt}{\emph{w.r.t.}\xspace}
\newcommand{\cf}{\emph{cf.}\xspace}
\newcommand{\etal}{et al.\xspace}
\newcommand{\aka}{\emph{a.k.a.},\xspace}

\newcommand{\taxounified}{\textit{T\textsubscript{unified}}}
\newcommand{\taxodspipelines}{\textit{T\textsubscript{dspipelines}}}
\newcommand{\taxodaswow}{\textit{T\textsubscript{daswow}}}

\newcommand{\datasetdspipelines}{\textit{D\textsubscript{dspipelines}}}
\newcommand{\datasetdaswow}{\textit{D\textsubscript{daswow}}}
\newcommand{\datasetrule}{\textit{D\textsubscript{rule}}}

\newcommand{\expert}[1]{\textit{E\textsubscript{#1}}}

\newcommand{\slmbest}{\textit{SLM\textsubscript{best}}}

\newcommand{\question}[1]{\textbf{RQ#1}}

\newcommand{\look}[1]{\colorbox{cyan}{#1}}

\newcommand{\Todo}[1]{\todo[inline]{#1}}
\newcommandx{\nl}[2][1=]{\todo[author=NL,inline, linecolor=green!40,backgroundcolor=green!40,bordercolor=green!40,#1]{#2}}
\newcommandx{\mbf}[2][1=]{\todo[author=MBF,inline, linecolor=red!70,backgroundcolor=pink!70,bordercolor=pink!70,#1]{#2}}
\newcommandx{\fred}[2][1=]{\todo[author=FP,inline, linecolor=yellow!40,backgroundcolor=yellow!40,bordercolor=yellow!40,#1]{#2}}
\newcommandx{\reviewer}[3][2=]{\todo[author=#1,inline, linecolor=red!40,backgroundcolor=red!40,bordercolor=red!40,#2]{#3}}


\tcbset{
  rockrq/.style={
    colback=white,
    colframe=black!10,
    boxrule=0.3mm,
    arc=1mm,
    outer arc=1mm,
    left=6pt,
    right=6pt,
    top=5pt,
    bottom=5pt,
    fonttitle=\bfseries,
    coltitle=black,
    toptitle=1mm,
    bottomtitle=1mm,
    title={#1}
  }
}

\newcommand{\researchquestion}[5]{
  \begin{tcolorbox}[rockrq, title={Research Question #1}]
    \label{rq:#1}
    \textbf{RQ#1:} #2

    \vspace{4pt}
    \textbf{Hypothesis (H\textsubscript{#1}):} #3

    \vspace{4pt}
    \textbf{Rationale:} #4

    \vspace{4pt}
    \textbf{Null Hypothesis (H\textsubscript{0.#1}):} #5
  \end{tcolorbox}
}


\newcolumntype{L}[1]{>{\RaggedRight\arraybackslash}m{#1}}

\definecolor{lightgray}{gray}{0.95}
\definecolor{lightergray}{gray}{0.98}

\title{Using Small Language Models to Reverse-Engineer Machine Learning Pipelines Structures}

\date{October 2025}

\author{
    \IEEEauthorblockN{Nicolas Lacroix\hypersetup{pdfborder={0 0 0}}\orcidlink{0009-0004-4390-6694}, Frédéric Precioso\hypersetup{pdfborder={0 0 0}}\orcidlink{0000-0001-8712-1443}}
    \IEEEauthorblockA{\textit{Université Côte d’Azur, Inria, CNRS, I3S}\\Sophia Antipolis, France\\firstname.name@univ-cotedazur.fr}
    \and
    \IEEEauthorblockN{Mireille Blay-Fornarino\hypersetup{pdfborder={0 0 0}}\orcidlink{0000-0001-9852-7764}}\IEEEauthorblockA{\textit{Université Côte d’Azur, CNRS, I3S}\\Sophia Antipolis, France\\firstname.name@univ-cotedazur.fr}
    \and 
    \IEEEauthorblockN{Sébastien Mosser\hypersetup{pdfborder={0 0 0}}\orcidlink{0000-0001-9769-216X}}\IEEEauthorblockA{\textit{McSCert, McMaster University}\\Hamilton, Ontario, Canada\\mossers@mcmaster.ca}
}

\maketitle

\begin{abstract}
    \textit{Background}: Extracting the stages that structure Machine Learning (ML) pipelines from source code is key for gaining a deeper understanding of data science practices. However, the diversity caused by the constant evolution of the ML ecosystem (\eg algorithms, libraries, datasets) makes this task challenging. Existing approaches either depend on non-scalable, manual labeling or on ML classifiers that do not properly support the diversity of the domain. These limitations highlight the need for more flexible and reliable solutions. \textit{Objective}: We evaluate whether Small Language Models (SLMs) can leverage their code understanding and classification abilities to address these limitations, and subsequently, how they can advance our understanding of data science practices. \textit{Method}: We conduct a confirmatory study based on two reference works selected for their relevance regarding current state-of-the-art’s limitations. First, we compare several SLMs using Cochran’s Q test. The best-performing model is then evaluated against the reference studies using two distinct McNemar’s tests. We further analyze how variations in taxonomy definitions affect performance through an additional Cochran’s Q test. Finally, a goodness-of-fit analysis is conducted using Pearson’s chi-squared tests to compare our insights on data science practices with those from prior studies.
\end{abstract}


\section{Introduction}
\label{s:intro}

Understanding how data scientists structure their Machine Learning (ML) pipelines is key to better supporting the quality and evolution of these pipelines. Studying them at an abstract level enables the analysis of how workflows are organized, iterated, and reused. This higher-level perspective is therefore crucial to characterize current practices empirically.

Recently, studies have examined pipeline quality~\cite{wang2025machine,islam2019comprehensive}, identified recurring patterns~\cite{bogner2021characterizing}, and supported practitioners through pipeline reuse~\cite{redyuk2024assisted}. All these efforts rely on accurately extracting the underlying structure of ML pipelines from source code. However, this task is often achieved manually because of the high diversity of the ML domain. Indeed, the exploratory nature~\cite{martinez2019crisp} of ML associated with the rapid evolution of new data processing algorithms and ML libraries has led to increasingly heterogeneous ML pipelines~\cite{biswas2022art,ramasamy2023workflow}.

From the reliance on static mappings between specific libraries and pipeline stages~\cite{jiang2022elevating,biswas2022art,venkatesh2023enhancing}, efforts have recently moved toward automated approaches with rule-based or supervised classifiers~\cite{wang2021graph,zhang2022coral,perez2024flexible}. Despite their promising results, they are still constrained in performance or in scope (\eg specific libraries or pipeline patterns). As a result, scale and reliability of current analyses of data science practices remain limited.

Simultaneously, language models have shown strong capabilities in code understanding and classification tasks~\cite{hou2024large}. In addition, Small Language Models (SLMs) have drawn interest from the scientific community, looking for more sustainable and reproducible practices. Yet, their potential for extracting ML pipeline structures has still to be evaluated.

In this confirmatory study, we will investigate whether SLMs can be used as a classification component in a reverse-engineering process to efficiently extract ML pipeline structures. Our contribution is threefold:

\begin{enumerate}
    \item An assessment of the performance gain of five SLMs, pre-selected based on their HumanEval performance~\cite{chen2021evaluating}, compared to existing works
    \item A statistical evaluation of the potential impact of taxonomy definition on classification results, using a mutation process 
    \item A goodness-of-fit (Pearson’s chi-squared) test to challenge the current understanding of data scientists' practices reported in existing literature
\end{enumerate}

Results of our work could open new perspectives about reliable ML pipelines characterization (\eg anti-pattern identification~\cite{garijo2013detecting,bogner2021characterizing}), ML transparency (\eg~\textit{data provenance} enrichment~\cite{namaki2020vamsa}), and data scientists' guidance (\eg visualizations and recommendation systems~\cite{redyuk2024assisted}).

\section{Background and Related Work}
\label{s:background_related_work}

\subsection{Reverse-Engineering and Language Models}
\label{ss:llm_reverse_engineering_tasks}

While code summarizing and documentation generation remain very common applications~\cite{hou2024large}, researchers have recently become interested in the ability of language models to extract structured information from code bases. More specifically, Boronat et al.~\cite{boronat2025mdre} have proposed a tool, \textit{MDRE-LLM}, that integrates LLMs with traditional Model-Driven Reverse Engineering (MDRE) techniques for extracting high-level abstractions from source code. Additionally, Siala et al.~\cite{siala2025comparison} studied the use of LLMs for reverse engineering Python and Java code to the OCL formal language. Similarly, Campanello et al.~\cite{campanello2025use} have shown promising results in the use of LLM (GPT-4) for reverse engineering class (as PlantUML) diagrams.

It is also crucial to adopt a rigorous methodology to properly support the probabilistic nature of language models. To that end, Wagner et al.~\cite{wagner2025towards} have proposed important preliminary guidelines. Most notably, open models, transparent prompts and models configuration are described as key factors in achieving reliable and reproducible results.

\subsection{Reverse-Engineering Applied to ML Pipelines Extraction}
\label{ss:auto_pipeline_extract}

Software engineering researchers have a long-standing interest in analyzing scientific workflows~\cite{stevens2001classification,wassink2009analysing} and managing computational processes~\cite{garijo2013detecting,liu2015survey,patterson2017dataflow}, with a growing focus on extracting ML pipeline structures from source code for understanding and evaluating practices.

Static mapping or rule-based approaches identify stages by matching code to predefined functions or libraries. For instance, initial work by Venkatesh et al.~\cite{venkatesh2023enhancing} identified stages based on a static mapping of 13 libraries, integrating these as documentation. Similarly, Jiang et al.~\cite{jiang2022elevating} proposed around 700 hard-coded rules derived from a few selected ML libraries and have shown that one cell can contain code instructions related to multiple stages. Biswas et al.~\cite{biswas2022art} used a static mapping composed of 403 specific functions to explore data scientists' practices, giving first insights about stages frequency and ordering in Kaggle notebooks.

Works leveraging supervised classifiers have tried to mitigate these static limitations. For example, Zhang et al.'s CORAL solution~\cite{zhang2022coral} uses an ML classifier trained on a custom taxonomy but is restricted to seven Data Science libraries (pandas, statsmodels, etc.). More recently, Venkatesh et al. has extended their initial work and proposed a supervised classifier~\cite{venkatesh2024static} trained on their previous mapping of 13 libraries. Similarly, Perez et al.~\cite{perez2024flexible} tried to improve this approach by combining a decision tree model with a rule-based model. In their comprehensive empirical study, Ramasamy et al.~\cite{ramasamy2023workflow} also developed a classifier and explored data scientists' practices. They confirmed the imbalanced stage frequencies and transition probabilities, but showed mixed results in their classifier evaluation (e.g., poor \textit{F}1 score of 0.29 for prediction), highlighting the need for more flexible reverse-engineering solutions.

Overall, these approaches have demonstrated promising results within specific contexts. However, their reliance on predefined rules or narrowly trained datasets limits their scalability and generalization to the highly variable ML ecosystem.

\section{Research Questions and Hypotheses}
\label{s:rqs_hypotheses}

Our confirmatory study targets two specific research goals:

 \begin{itemize}
     \item \textbf{Research Goal 1}: Evaluate the applicability of SLMs for extracting ML pipelines structures
     \item \textbf{Research Goal 2}: Challenge our current understanding of data scientists' practices
 \end{itemize}

The first goal will be reached by answering the two research questions \textbf{RQ1} and \textbf{RQ2} while answering \textbf{RQ3} will contribute to the second goal.

\researchquestion
  {1}
  {In ML pipeline's structure extraction, are SLMs significantly more accurate in identifying stages than existing model-based or rule-based classifiers?}
  {SLMs are more accurate than existing model-based or rule-based classifiers to extract ML pipelines' structures (stages) from ML code.}
  {As SLMs have been trained on large code datasets and documentation about Data Science lifecycle, they should better support the high diversity of the ML domain to extract ML pipelines' structures.}
  {SLMs are \textbf{not} more accurate than existing model-based or rule-based classifiers to extract ML pipelines' structures (stages) from ML code.}

\researchquestion
  {2}
  {Does the taxonomy ``wording'' significantly influence the classification results of SLMs?}
  {The words used in the taxonomy impacts the classification results of SLMs.}
  {When using language models, each word corresponds to one or more tokens, impacting the probabilities of the next token to be generated. As a consequence, changing the words used in the taxonomy (\eg ``Modeling`` to ``Model Building`` should have an impact on the results.}
  {The choice of words does \textbf{not} significantly impact the classification results.}

\researchquestion
  {3}
  {Does the use of an SLM in the reverse-engineering process significantly affect our understanding of data scientists' practices, compared to reference studies?}
  {Using an SLM to extract ML pipelines structures significantly affects our understanding of data scientists' practices, compared to existing studies.}
  {Using an SLM as a classifier that better supports the diversity and heterogeneity of ML code should reveal different patterns in ML pipelines than those reported by existing exploration studies~\cite{biswas2022art,ramasamy2023workflow}.}
  {Using an SLM does \textbf{not} significantly affect our understanding of data scientists' practices, compared to existing studies.}

\section{Variables}
\label{s:variables}

We introduce the variables of our study in \cref{tab:variables}. To answer \textbf{RQ1}, we manipulate the classification model and measure the classification performance, the duration and the SLM perplexity. We also manipulate the taxonomy as part of \textbf{RQ2} and measure the same dependent variables. Finally, \textbf{RQ3} is measured by reproducing insights reported by reference studies. Being mainly related to the use of SLMs, confounding variables effects are controlled by reusing existing datasets and ensuring a common language model configuration.

\begin{table*}[t]
\centering
\caption{Variables of the confirmatory study}
\label{tab:variables}
\vspace{0.5em}

\textbf{Independent variables}
\vspace{0.3em}

\begin{tabular}{L{2.5cm} L{1.5cm} L{5cm} L{1.2cm} L{5cm}}
\toprule
\textbf{Name} & \textbf{Abbreviation} & \textbf{Description} & \textbf{Scale} & \textbf{Operationalization} \\
\midrule
\rowcolor{lightgray}
Classification Model & \textit{M\textsubscript{classification}} & The model used for classifying ML pipelines code: Rule-based, Supervised or SLM classifier & Nominal & see \ref{ss:reference_studies} and \ref{ss:slm_tools_resources} for models selection \\
Taxonomy & \taxounified & The taxonomy being used to define all possible stages in an ML pipeline & Nominal & Unique taxonomy \taxounified~(\cf\ref{ss:taxonomies_unification}) for RQ1 and RQ3 and mutated versions for RQ2 (\cf\ref{ss:taxonomy_mutation}) \\
\bottomrule
\end{tabular}

\vspace{0.3em}

\textbf{Dependent variables}
\vspace{0.3em}

\begin{tabular}{L{2.5cm} L{1.5cm} L{5cm} L{1.2cm} L{5cm}}
\toprule
\textbf{Name} & \textbf{Abbreviation} & \textbf{Description} & \textbf{Scale} & \textbf{Operationalization} \\
\midrule
\rowcolor{lightgray}
Classification Performance & \textit{C\textsubscript{performance}} & the performance of the classification, including a per-class evaluation & Ratio & Matthews Correlation Coefficient (MCC) to better support unbalanced classes and F1-score for each class. \\ 
Classification duration & \textit{C\textsubscript{duration}} & the duration of the classification (\ie inference duration) & Ratio & Duration of the classification (from first to last token for the SLM) in milliseconds \\
\rowcolor{lightgray}
SLM perplexity & \textit{SLM\textsubscript{perplexity}} & (only for the SLM) the perplexity value of the SLM over the classified code & Ratio & The average perplexity of generated tokens \\
Exploration insights & \textit{E\textsubscript{insights}} & the insights gathered from the analysis of extracted ML pipeline structures & Ratio & Reproduction of insights reported by reference studies \\
\bottomrule
\end{tabular}

\vspace{0.3em}

\textbf{Confounding variables}
\vspace{0.3em}

\begin{tabular}{L{2.5cm} L{1.5cm} L{5cm} L{1.2cm} L{5cm}}
\toprule
\textbf{Name} & \textbf{Abbreviation} & \textbf{Description} & \textbf{Scale} & \textbf{Operationalization} \\
\midrule
\rowcolor{lightgray}
Notebooks Diversity & \textit{N\textsubscript{diversity}} & The diversity of the notebooks used for evaluation & Nominal & We will reuse reference studies' datasets to have the same notebooks diversity \\
SLM inference determinism & \textit{SLM\textsubscript{determinism}} & The inference of SLMs is not deterministic by default, leading to a lack of reproducibility of the results & Binary & We will use a Docker image making the vLLM inference server deterministic~\cite{he2025nondeterminism} and follow the guidelines established by OpenAI for classification tasks \\
\rowcolor{lightgray}
SLM specialization & \textit{SLM\textsubscript{specialization}} & The ``specialization'' (generalist vs code) of the Large Language Model being used for ML pipelines structures extraction & Nominal & As we aim at classifying code, we will focus on SLM specifically trained for code \\
SLM size & \textit{SLM\textsubscript{size}} & The size of the selected SLMs & Interval & See~\ref{sss:resources_reproducibility} for SLMs selection \\
\rowcolor{lightgray}
Prompt content & \textit{P\textsubscript{content}} & The content of the prompt affects the predictions of the SLM & Nominal & We will use the same base prompt for all SLMs \\
Prompting technique & \textit{P\textsubscript{technique}} & The advanced prompting technique used to improve the classification process & Nominal & We will use the same prompting technique (\eg Chain-Of-Thought, Few-Shot) for all SLMs \\
\bottomrule
\end{tabular}
\end{table*}

\section{Participants/Subjects/Datasets}
\label{s:participants_subjects_datasets}

\subsection{Reference Studies}
\label{ss:reference_studies}

Reference studies were selected according to the following criteria:

\begin{enumerate}
    \item \textit{C\textsubscript{1}}: the study reverse-engineers ML pipeline structures (ML stages) from code
    \item \textit{C\textsubscript{2}}: the study presents empirical exploration results on data scientist practices, necessary for comparison in \textbf{RQ3}
    \item \textit{C\textsubscript{3}}: the study provides an executable, open-source reproduction package to ensure reproducibility.
\end{enumerate}

After selection of works matching \textit{C\textsubscript{1}} from the literature, we filtered out works either violating \textit{C\textsubscript{2}} (\cite{wang2022documentation,wang2021graph,patterson2017dataflow,zhang2022coral,garijo2013detecting}) or \textit{C\textsubscript{3}} (\cite{perez2024flexible,venkatesh2024static,venkatesh2023enhancing}). This left us with three studies \cite{ramasamy2023workflow,biswas2022art,jiang2022elevating} that we compared based on their \textit{depth} (single instruction vs. entire notebook cell classification), \textit{breadth} (library coverage), and \textit{publication date}. The study \textit{DS-Pipelines} stood out as the only one to classify code at the instruction level, while \textit{DASWOW} was the most recent study to avoid explicitly relying on a subset of libraries.

As a consequence, these two noteworthy studies form the reference baselines for our confirmatory study:

\begin{itemize}
  \item \textit{DS-Pipelines\footnote{\url{https://github.com/sumonbis/DS-Pipeline}}}~\cite{biswas2022art}: works at the instruction level and relies on a static mapping (rule-based classifier), created manually.
  \item \textit{DASWOW\footnote{\url{https://zenodo.org/records/7109939}}}~\cite{ramasamy2023workflow}: works at the cell level and relies on a supervised classifier (Random Forest Classifier)
\end{itemize}

We were able to reproduce the results of both studies from their reproduction packages, confirming our ability to conduct our study.

\subsection{Datasets and Taxonomy}
\label{ss:datasets}

\subsubsection{Original datasets for comparison}
\label{sss:original_datasets_comparison}

We selected the original datasets \datasetdspipelines~and \datasetdaswow~from the two reference studies (\cf~\cref{ss:reference_studies}) to serve as the ground truth for our comparative analysis.

\begin{itemize}
  \item \datasetdspipelines~consists of a set of 105 notebooks and a static mapping file \textit{stages.csv}. This file maps 404 functions or classes names to a stage (encoded as a number). The result dataset is obtained by applying this mapping to each notebook instruction using an open-source parser.
  \item \datasetdaswow~consists of 470 notebooks (9,678 code cells) organized in 3 dataframe: \textit{test\_features} (1918x32), \textit{train\_features} (5833x32), \textit{validation\_features} (1927x32). For each row, corresponding to a cell of a notebook, columns corresponding to the steps in the taxonomy have the value 0 or 1. Additionally, some metadata such as the filename, cell\_type, text (code content), etc. are present. As we intend to label each instruction, we will apply the AST-based parser of \textit{DS-Pipelines} to each notebook. However, code indentation has been lost during the dataframes construction by the \textit{DASWOW} authors, making it impossible to properly retrieve the corresponding AST. Therefore, we will retrieve each cell content from the original dataset \datasetrule\footnote{\url{https://library.ucsd.edu/dc/object/bb2733859v}} created by Rule et al.~\cite{rule2018exploration}.
\end{itemize}

\subsubsection{Taxonomy}
\label{sss:taxonomy}

The reference studies employ two distinct taxonomies:

\begin{itemize}
    \item \textit{DS-Pipelines} introduces a taxonomy~\taxodspipelines~of 11 stages, with only 6 of them being used when analysing notebooks (``\textit{in-the-small}'' analysis)
    \item \textit{DASWOW} introduces a taxonomy~\taxodaswow~of 10 stages derived from 4 studies~\cite{zhang2022coral,garijo2013detecting,wang2022documentation,aggarwal2019can}.
\end{itemize}

Our study will be based on a unified taxonomy~\taxounified, derived from the two reference taxonomies. The unification process is described in the execution plan in~\cref{ss:taxonomies_unification}. Each stage will be associated with a definition to comply with current practices and facilitate its reuse in future work. As we will prompt the SLMs using~\taxounified, we will need to operationalize the unification mapping to compare our results with the reference studies using \taxodspipelines~and \taxodaswow~(\cf~\cref{ss:comparison_reference_studies}).

\subsubsection{Power Analysis}
\label{sss:sampling_power_analysis}

Before proceeding with our study, we will conduct an \textit{a priori} power analysis to determine if the dataset size (\ie minimum between \datasetdspipelines~and \datasetdaswow~sizes) limits the risk for \textit{Type I} and \textit{Type II} errors. We will use the \textit{G*Power} software with $\alpha=0.05$ and $1-\beta=0.8$ (power) for the non-parametric tests for paired data (McNemar, Cochran's Q) as explained in the literature~\cite{card2020little}. Same values will be used for the goodness-of-fit test as recommended by Berman et al.~\cite{berman2016essential}. In case of insufficient dataset size, we will extend it with data from \datasetrule~and experts annotation (\cf\cref{ss:experts}).

\subsection{Experts}
\label{ss:experts}

We are working with 3 expert data scientists (\expert{1}, \expert{2}, \expert{3}) from academia, selected for their extensive experience related to Data Science practices. They will assist us in the elaboration of \taxounified~(\cf~\cref{ss:taxonomies_unification}), the assessment of \datasetdspipelines~and \datasetdaswow~quality and serve as independent annotators for any evaluation dataset extension (\cf\cref{sss:sampling_power_analysis}). For each task, they will be provided with the two reference studies to gain a proper understanding of the context.

\subsection{Small Language Models, Resources and Tools}
\label{ss:slm_tools_resources}

\subsubsection{Models Openness}
\label{sss:models_openness}

To promote openness in SLMs and ensure the reproducibility of our experiments, we only consider open-source and open-weight models. Among these models, we focus on those that have been specifically trained for code-related tasks. Being the \textit{de facto} standard, the selected hosting platform for these models is HuggingFace.

\subsubsection{Resources and Reproducibility}
\label{sss:resources_reproducibility}

We also consider computing resources and energy consumption as important factors that can limit the reproducibility of our experiments. Indeed, it constitutes a ``Barriers in the culture of research'' as described in the ``Sources of Non-Reproducibility'' by the National Academies of Sciences, Engineering, and Medicine~\cite{national2019reproducibility}. Furthermore, we aim to contribute to greater environmental sustainability, a core principle in ``The Copenhagen Manifesto''~\cite{russo2024generative}, by minimizing the environmental impact of our study. Accordingly, we consider SLMs to be language models with less than 8 billions parameters.

\subsubsection{Benchmark for SLMs Selection}
\label{sss:benchmark_slms_selection}

After exclusion of incompatible models based on the criteria discussed in previous sections, the selected models will be the 5 best performing models on the HumanEval~\cite{chen2021evaluating} benchmark at the moment of the realization of the experiment. The choice of the benchmark is motivated by its maturity and relevance for code-related tasks.

\subsubsection{Tools}
\label{sss:tools}

The core inference server used for the experiments will be vLLM (v0.10.2), a reference open source solution supporting HuggingFace-hosted models and Docker deployment. Prompt-engineering and monitoring will be done using a local instance of the open source project Langfuse (v3.114.0)\footnote{~\url{https://langfuse.com/docs}}. Insights from the reverse-engineering of ML pipelines will be collected using~\textit{Colombus}. This platform was developed in preparation for the experiment, tested with manually labeled data and will be made open source along with the exploration results. Finally, the statistical analysis will be achieved using dedicated Python and R scripts and open source scientific libraries, all included in the reproduction package.

\section{Execution Plan}
\label{s:execution_plan}

\subsection{Taxonomies Unification}
\label{ss:taxonomies_unification}

The first step in our execution plan is to construct the unified taxonomy~\taxounified~by asking experts to create a mapping $y$, using the power set $\mathcal{P}$, between the two reference taxonomies~\taxodspipelines~and \taxodaswow~(\cf~\cref{sss:taxonomy}).

{\setlength{\abovedisplayskip}{0pt}
\[
\small
\begin{array}{c}
f : \taxodspipelines \to \mathcal{P}(\taxodaswow) \\[2pt]
g : \taxodaswow \to \mathcal{P}(\taxodspipelines) \\[2pt]
y = \{\, (a, b) \in \taxodspipelines \times \taxodaswow \;\Big|\; a \in g(b) \lor b \in f(a) \,\} \\[4pt]
\end{array}
\]
}

The mapping $y$ is then merged by identifying connected components ($\text{CC}$), which are used to form groups of equivalent terms from the two taxonomies. A function $w$ then associates a headword to each group to obtain~\taxounified.

{\setlength{\abovedisplayskip}{0pt}
\[
\small
\begin{array}{c}
y' = \left\{\, 
(C \cap T_a,\; C \cap T_b) \;\middle|\; C \in \text{CC}(T_a \cup T_b, y) 
\,\right\} \\
\taxounified = \{\, w\bigl(A_C, B_C\bigr) \;\Big|\; (A_C, B_C) \in y' \,\}
\end{array}
\]
}

Unmatched stages will be kept or removed based on their relevance in the exploration analysis and corresponding decisions will be recorded. Having 3 experts working on all stages of the taxonomies, the inter-annotator agreement will be measured using Fleiss' Kappa $\kappa$. Potential disagreements will be solved using the majority vote.

\subsection{SLMs Configuration}
\label{ss:slm_configuration}

We will control SLM-related confounding variables (\cf\cref{tab:variables}) by defining a common configuration for all SLMs used in our experiment. This is crucial as inference parameters and prompting techniques can significantly impact classification output. First, we will set $\textit{temperature}=0$ and $\textit{top\_p}=1$ (nucleus sampling) to limit predictions randomness. Then, the \textit{max\_tokens} parameter will be adjusted to match the maximum length of the taxonomy's headwords (\cf\cref{ss:taxonomies_unification}). Finally, we will iterate over different prompting techniques as detailed by Schulhoff et al.~\cite{schulhoff2024prompt} to select the best performing one. Each prompt efficiency will be evaluated using the first selected SLM against the \textit{test features} part of \datasetdaswow~and evaluation results will be saved in the reproduction package.

\subsection{Comparison With Reference Studies}
\label{ss:comparison_reference_studies}

The comparison with reference studies will be achieved using the best performing SLM, \slmbest~(\cf\cref{ss:best_slms_selection_analysis}). We will extract every notebook's code and parse it using an AST-based parser. We will prompt the SLM for each yielded code instructions with the full notebook code in the context. To facilitate this comparison with the original ground truth, \slmbest~predictions on \taxounified~will be mapped back to the original taxonomies (\taxodaswow~and \taxodspipelines) using an operationalization of $y'$ defined in~\cref{ss:taxonomies_unification}.

\paragraph{Comparison with \textit{DS-Pipelines} (Instruction-Level)} Because both \textit{DS-Pipelines} and \slmbest~classify code at the instruction level, the comparison is straightforward as we will compare each identified instruction's stage with \textit{DS-Pipelines} equivalent.

\paragraph{Comparison with \textit{DASWOW} (Cell-Level and Multi-Label)} The comparison with \textit{DASWOW} is more complex as it classifies each notebook cell (rather than single code instructions) and potentially assigns multiple stages per cell. To enable comparison with our solution, we will classify each instruction using \slmbest, group them by respective cell and collect the result set of stages. Then, we will compare each cell's set of stages with the \textit{DASWOW} equivalent.

The statistical tests are detailed in the corresponding analysis~\cref{ss:reference_studies_comparison_analysis}.

\subsection{Taxonomy Mutation}
\label{ss:taxonomy_mutation}

The taxonomy mutation aims to evaluate the robustness of the SLM's classification performance against linguistic (potentially semantic) variations in the stage definitions (\ie headwords). This mutation process consists of creating different versions of \taxounified~by replacing a headword with a synonym. We will repeat this operation a maximum of 3 times for each headword until a significant impact is identified (\cref{ss:taxonomy_words_analysis}).

\subsection{Data Scientists Practices Exploration}
\label{ss:datascientists_practices_exploration}
The final part of our study will involve an exploration platform, \textit{Colombus} (\cf\cref{sss:tools}). It features a visualization interface (\eg visualize common stages patterns) and a querying component (\eg query specific patterns, get metrics for each stage). Insights reported by reference studies (\eg stages frequency, transition probabilities, distribution of cell and instructions for each stage) will be replicated with results from \slmbest. Additional insights and visualizations (\eg heat map of frequent patterns) will contribute to a greater understanding of data scientists practices. The corresponding analysis is detailed in~\cref{ss:practices_exploration_analysis}.

\section{Analysis Plan}
\label{s:analysis_plan}

\subsection{H1. SLMs are more accurate than existing model-based or rule-based classifiers to extract ML pipelines structures (stages) from ML code}
\label{ss:TODO}

\subsubsection{Best Performing SLM Analysis}
\label{ss:best_slms_selection_analysis}

Given the reported unbalanced stage frequencies in the two reference studies, we will evaluate~\textit{C\textsubscript{performance}} for each SLM using the Matthews Correlation Coefficient (MCC). The comparison of the obtained coefficients will permit the identification of a best performing SLM,~\slmbest. The heterogeneity in the results will be statistically evaluated using the Cochran's Q test with predictions converted to a binary outcome (valid or invalid prediction). If any significant effect is detected, we will do a post-hoc analysis with multiple McNemar's tests to properly identify these differences.

Finally, to better understand strengths and weaknesses of \slmbest, including differences hidden by the previous binary transformation (\eg imbalance of performance on infrequent classes), we will report:

\begin{itemize}
    \item the overall accuracy and the f1-score value for each stage to highlight potential performance gap between stages
    \item the confusion matrix to better understand limitations of \slmbest~considering \taxounified
    \item the SLM perplexity to properly quantify the uncertainty of the predictions
\end{itemize}

\subsubsection{Comparison With Reference Studies}
\label{ss:reference_studies_comparison_analysis}

To reject the null hypothesis~\textbf{H\textsubscript{0.1}}, we must demonstrate that \slmbest~performs significantly better than existing approaches. Therefore, we will first assess the significance of the performance differences using two separate McNemar's tests with predictions converted to a binary outcome (valid, invalid prediction). The first test will involve \textit{DASWOW} and \slmbest~on \datasetdaswow. The second test will involve \textit{DS-Pipelines} and \slmbest~on \datasetdspipelines. Secondly, we will calculate the MCC score for each reference study and compare it with \slmbest~to identify if our approach performs better than the existing studies.

If both statistical tests show a significant difference and the MCC scores show that \slmbest~performed better than the reference study, we will be able to claim that \slmbest~is effectively more accurate than model-based or rule-based classifiers to extract ML pipelines structures (stages) from ML code.

For a more comprehensive analysis, we will also compare the f1-score of each stage to properly understand where each approach performs the best. Finally, we will compare the duration of the inference in order to put the potential performance gain into perspective in relation to its cost.

\subsection{H2. The words used in the taxonomy impacts the classification results of SLMs}
\label{ss:taxonomy_words_analysis}

To reject the null hypothesis \textbf{H\textsubscript{0.2}}, we must show that there is a significant difference in the classification results when using \slmbest~on \taxounified~and the mutated taxonomies as detailed in~\cref{ss:taxonomy_mutation}.

We will use the Cochran's Q test, similarly to~\cref{ss:best_slms_selection_analysis}, with each treatment being \slmbest~prompted with a different version of the taxonomy. If a statistically significant difference is detected across the treatments, we will be able to state that words used in the taxonomy impacts the classification results of SLMs.

Finally, we will report the same detailed measures as in \cref{ss:best_slms_selection_analysis}  for each significant variation to better inform about the impact of changes made to the taxonomy wording.

\subsection{H3. Using an SLM to extract ML pipelines structures permits the identification of different practices compared to existing studies}
\label{ss:practices_exploration_analysis}

To reject the null hypothesis~\textbf{H\textsubscript{0.3}}, we must show that reported exploration insights from reference studies significantly differ from our study. Based on collected insights (\cf\cref{ss:datascientists_practices_exploration}), we will use the Pearson’s chi-squared test to realize a goodness-of-fit test where results from \slmbest~are the observed frequencies and the other studies' results are the expected frequencies. When the insights are reported as a matrix (\eg transition matrix), we will flatten it and use the same test for comparison.

However, some insights cannot be compared using the previously discussed test (\eg representative data science pipeline from \textit{DS-Pipelines}) as they are descriptive rather than distributional. Since these insights are generally the result of previously compared quantitative data, we will report and discuss them qualitatively but will not include them as a criterion for the formal rejection of the null hypothesis.

\section{Threats to Validity}
\label{s:threats_validity}

\subsection{Internal Validity}
\label{ss:internal_validity}

Reusing datasets from the two peer-reviewed reference studies may pose a threat as we do not control their diversity (\eg libaries) and quality. However, this choice avoids adding new biases from new annotators and does not prevent us from measuring this diversity. Additionally, our experts will assess their quality before conducting the experiments. Another threat comes from the potential contamination with~\datasetdspipelines~or~\datasetdaswow~being present in the training datasets of selected SLMs. As we do not necessarily have access to the original training datasets, we will not be able to confidently determine whether contamination has occurred. Yet, the design of our study has the advantage of using a unique taxonomy, unknown to the model. Furthermore, the risk is limited as we will prompt \slmbest~with one raw line of code at a time while \datasetdaswow~contains the mapping for each cell and \textit{DS-Pipelines} limits the mapping to the function been called with the stage. While a similar limitation related to contamination also exists with the HumanEval benchmark, this choice has been made in accordance with the most recent recommendations~\cite{wagner2025towards}. Lastly, the prompting technique selection has been done using only one of the SLMs. This may give it an advantage over the other models (\cf\cref{ss:best_slms_selection_analysis}).

\subsection{External Validity}
\label{ss:external_validity}

The selection of SLMs specifically trained for code-related tasks seems reasonable as we study ML code. However, further studies are needed to evaluate the interest of such restriction. Furthermore, using language models involve having a knowledge cutoff that can lead to a decrease of performance in the future. This phenomenon should be properly evaluated in the next few years. Additionally, our analysis relies on two studies which may limit the generalizability of our conclusions to the full diversity of ML domain. However, the selection process involves an evaluation of their relevancy and the detailed methodology with reproduction package should facilitate complementary empirical studies. Finally, the decision to limit reference studies to those offering accessible reproduction packages may prevent us from finding more effective solutions, but it promotes open science and ensures our comparisons are based on reproducible artifacts.

\subsection{Construct Validity}
\label{ss:construct_validity}

One threat to construct validity regarding the evaluation of classification performance comes from the unbalanced nature of stages. For instance, the data preparation stage represents a more extensive part of the ML pipeline than model evaluation. We mitigate this threat by using the Matthews Correlation Coefficient (MCC) which supports well unbalanced classes in classification tasks. We supplement this analysis with the f1-score for each stage to measure the performance on each stage independently. The broad scope of the ``exploration insights'' construct, encompassing the diverse practices of data scientists during ML pipeline development, also poses a threat. We mitigate this by relying on reported insights from the reference studies as reflective indicators to measure it. However, reporting additional insights could deepen the analysis. Therefore, we will present new forms of insight (\eg heat maps of frequent patterns) to enrich the understanding of the construct.

\section{Acknowledgments}
\label{s:acknowledgments}

This project has received funding from the French ANR Under Grant Agreement No. ANR-24-IAS2-0002-01 (TSIA 2024 – Specific Topics in Artificial Intelligence) and ANR-24-CE25-1286-01.

\bibliographystyle{plain}
\bibliography{references}

\end{document}